# A proposal for a coordinated effort for the determination of brainwide neuroanatomical connectivity in model organisms at a mesoscopic scale

Jason W. Bohland[1], Caizhi Wu, Helen Barbas[2], Hemant Bokil, Mihail Bota[3], Hans C. Breiter[4], Hollis T. Cline, John C. Doyle[5], Peter J. Freed[6], Ralph J. Greenspan[7], Suzanne N. Haber[8], Michael Hawrylycz[9], Daniel G. Herrera[10], Claus C. Hilgetag[11], Z. Josh Huang, Allan Jones, Edward G. Jones[12], Harvey J. Karten[13], David Kleinfeld[14], Rolf Kötter[15], Henry A. Lester[16], John M. Lin, Brett D. Mensh[17], Shawn Mikula, Jaak Panksepp[18], Joseph L. Price[19], Joseph Safdieh[20], Clifford B. Saper[21], Nicholas D. Schiff, Jeremy D. Schmahmann[22], Bruce W. Stillman, Karel Svoboda[23], Larry W. Swanson, Arthur W. Toga[24], David C. Van Essen, James D. Watson, Partha P. Mitra[1]

[1] Cold Spring Harbor Laboratory, 1 Bungtown Rd, Cold Spring Harbor, NY 11724
[2] Department of Health Sciences, Boston University, 635 Commonwealth Ave, Boston, MA 02215
[3] Department of Biological Sciences, University of Southern California, Los Angeles, CA 90089
[4] Department of Radiology, Massachusetts General Hospital, 149 Thirteenth Street, Charlestown, MA 12129
[5] Department of Electrical Engineering, California Institute of Technology, Pasadena, CA 91124
[6] NYS Psychiatric Institute, Columbia University Medical Center, 1051 Riverside Drive, New York, NY 10032
[7] The Neurosciences Institute, 10640 John Jay Hopkins Drive, San Diego, CA 92121
[8] Department of Pharmacology & Physiology, University of Rochester Medical Center, Rochester, NY 14642
[9] Allen Institute for Brain Science, 551 North 34th Street, Suite 200, Seattle, WA 98103
[10] Department of Psychiatry, Weill Cornell Medical College, New York, NY 10021
[11] School of Engineering and Science, Jacobs University Bremen, 28725 Bremen, Germany
[12] Center for Neuroscience, University of California at Davis, 1544 Newton Court, Davis, CA 95616
[13] Department of Neurosciences, University of California San Diego School of Medicine, La Jolla, CA 92093
[14] Department of Physics, University of California at San Diego, La Jolla, CA 92093
[15] Department of Cognitive Neuroscience, Radboud University, Nijmegen, NL
[16] Department of Biology, California Institute of Technology, Pasadena, CA 91125
[17] Department of Psychiatry, Columbia University Medical Center, 1051 Riverside Drive, New York, NY 10032
[18] College of Veterinary Medicine, PO Box 646520, Washington State University, Pullman, WA 99164
[19] Department of Anatomy & Neurobiology, Washington University School of Medicine, St. Louis, MO 63130
[20] Department of Neurology, Weill Cornell Medical College, New York, NY 10021
[21] Department of Neurology, Beth Israel Deaconess Medical Center, 330 Brookline Ave, Boston, MA 02215
[22] Department of Neurology, Massachusetts General Hospital, Charles River Plaza, South Building, Boston, MA 02114
[23] Janelia Farm Research Campus, Howard Hughes Medical Institute, Ashburn, VA 20147
[24] Laboratory of NeuroImaging, Department of Neurology, UCLA School of Medicine, Los Angeles, CA 90095

*Abstract*

In this era of complete genomes, our knowledge of neuroanatomical circuitry remains surprisingly sparse. Such knowledge is however critical both for basic and clinical research into brain function. Here we advocate for a concerted effort to fill this gap, through systematic, experimental mapping of neural circuits at a mesoscopic scale of resolution suitable for comprehensive, brain-wide coverage, using injections of tracers or viral vectors. We detail the scientific and medical rationale and briefly review existing knowledge and experimental techniques. We define a set of desiderata, including brain-wide coverage; validated and extensible experimental techniques suitable for standardization and automation; centralized, open access data repository; compatibility with existing resources, and tractability with current informatics technology. We discuss a hypothetical but tractable plan for mouse, additional efforts for the macaque, and technique development for human. We estimate that the mouse connectivity project could be completed within five years with a comparatively modest budget.

*Introduction*

The defining architectural feature of the nervous system is that it forms a circuit. Unlike other tissues or organs, it is the patterns of axonal connections between neurons that determine the functioning of the brain. Nevertheless, more than a decade after Francis Crick and Ted Jones bemoaned the "Backwardness of Human Neuroanatomy [1]," our empirical knowledge about neuroanatomical connectivity in model organisms, including the mammalian species most widely used in biomedical research, remains surprisingly sparse. Efforts to manually curate neuroanatomical knowledge from the literature currently provide information about the reported presence or absence of ~10% of all possible long-range projections between the roughly 500 identified brain regions in the rat [2] (Figure 1). While this number does not represent a comprehensive survey of the literature, it is clear that many possible projections have not yet been studied using modern tracing methods. In addition, the standard level of data analysis and presentation provides only a qualitative view of the known projections. Such paucity of empirical knowledge stands in contrast with the complete genomes now available for many organisms.

Here we argue the case for a coordinated effort across the neuroscience community to comprehensively determine neuroanatomical connectivity at a brainwide level in model organisms including the mouse, macaque and eventually human. We discuss the important issues of resolution and rationale and survey the state of current knowledge and available techniques, then offer a basic outline for an experimental program and associated informatics requirements. The Allen Brain Atlas (ABA) for gene expression in the mouse [3] has demonstrated both the power of scaling up standardized techniques in neuroanatomical research and the feasibility of brainwide approaches. Numerous follow-up efforts to genome projects are also under way at various levels leading up to the phenotype. Time is therefore ripe for brainwide connectivity projects, to modernize neuroanatomical research and to fill perhaps the largest lacuna in our knowledge about nervous system structure. The purpose of this article, which has resulted from discussions with a large and varied working group of experts, is to provide motivation and background for readers interested in brainwide connectivity projects, estimate resource requirements by analyzing a feasible scenario, recommend directions for such projects and provide a platform for further discussions. The issues discussed here are likely to be relevant in implementing such a project through any combination of centralized and distributed efforts.

*The mesoscopic level of resolution*

It is clear that there exists some degree of non-random organization of the interconnections in the nervous system at multiple scales including individual neurons, groups of neurons, architectonic regions and subcortical nuclei, and functional systems. *Macroscopic* brain organization, at the level of entire structural-functional systems and major fiber bundles is somewhat understood but provides an insufficient description of the overall architecture. However, for complex vertebrate brains it is not currently technologically feasible to determine brainwide connectivity at the level of individual synapses. Further, while a statistical description is possible at this *microscopic* resolution, correspondence cannot be expected between individual brains described at the level of all synapses of all neurons. Significantly more invariance can be expected at a *mesoscopic* level where co-localized groups of neurons, perhaps of the same type or sharing common organizational features, are considered together as a unit, and projection patterns from these neuronal groups are studied over macroscopic distances. This level of connectivity is well-suited to aid our understanding of specific mental functions. A comprehensive mesoscopic wiring diagram, if available, would supply a meaningful skeleton that can be further augmented by the statistical characterization of microcircuitry at a finer scale (e.g. single neurons or cortical columns).

The existence and nature of invariant connectivity patterns across individual brains is itself a topic of research which can be addressed within a large-scale connectivity project. There is adequate evidence for mesoscopic architectural invariance in the form of cyto-, chemo-, and myelo-architectonically defined brain regions and from spatial gene expression patterns to proceed. In addition, however, a brainwide project executed with calculated redundancy will make it possible to empirically define the extent of such invariance. Further, if input and output connections are methodically determined along an appropriate anatomical grid it should be possible to delineate the mesoscopic projection patterns in brain space without imposing a system of discrete anatomical parcels defined *a priori*.

*Scientific rationale*
The availability of mesoscopic circuit diagrams for model nervous systems would impact neuroscience research at nearly all levels. Because connectivity underlies nervous system function, any lack of such knowledge impedes the achievement of comprehensive understanding, even if complete information was present about cytoarchitecture, neuronal cell types, gene expression profiles, or other structural considerations. Furthermore, the connectional architecture of the nervous system – the *connectivity phenotype* – is a critical missing *link between genotype and behavioral phenotype*; the simultaneous availability of comprehensive genomic and neuroanatomical information will greatly narrow this gap. The scientific rationale can be further sharpened by examining the role of circuitry in experimental and theoretical approaches to the nervous system.

*Experimental design* in electrophysiological studies can be improved by explicit consideration of connectivity. For example, without any reference to underlying connectivity it is difficult to interpret measured physiological activity or the effects of microstimulation. Studies that consider the internal dynamics of the brain, including studies of selective attention, often make arguments about top-down or bottom-up processes, which are ultimately contingent on neuroanatomical information that is frequently deficient. Likewise, the lack of *empirical constraints on neural network models* remains an Achilles heel of that subject area, and such theoretical research would benefit greatly from added knowledge of connectional brain architecture.

Many *comparative and evolutionary studies* have also suffered from a phrenological emphasis on changes in morphological characteristics and relative sizes of parts of the nervous system, with less consideration of connectivity. Knowledge of the mesoscopic circuit diagrams for multiple model organisms will greatly advance comparative and evolutionary neuroanatomy, as has been the case for comparative and evolutionary genomics. This is highlighted by recent advances in understanding the relation between avian and mammalian brains. Purely structural considerations, such as the presence of a layered cortex in mammals, had led to incorrect homological identification of avian telencephalic structures with mammalian basal ganglia.

Connectivity considerations have led to a profound revision of this view, leading to a new nomenclature for avian brain compartments [4].

*Biomedical rationale*
Neurological and neuropsychiatric disorders are responsible for approximately 30% of the total burden of illness in the United States according to the World Health Organization's estimated Disability Adjusted Life Years (DALYs) for 2002 (http://www.who.int/healthinfo/bodgbd2002revised/en/). The dominant paradigms for understanding such disorders have involved focal lesions, widespread neurodegeneration, vascular compromise, and neurotransmitter dysregulation, with circuit considerations playing a comparatively minor role. It has long been known, however, that disruptions in neural connectivity can underlie human brain disease [5,6]. In disorders with no identified genetic component (e.g. traumatic brain injury or infectious disease), dysfunction arises directly from a disruption of the normal circuit. For those with heritable susceptibility effects, genetic polymorphism and cellular processes play a greater role, but anatomical circuits remain critical to understanding symptoms and developing therapies. In Parkinson's disease, for example, drug and stimulation-based therapeutic interventions do not occur at the cellular lesion site, but rather are contingent on understanding interactions within the extra-pyramidal motor system [7]. Incomplete knowledge of this circuitry potentially holds back development of therapies for both Parkinson's and Huntington's diseases, despite a reasonably complete understanding of the genetic etiology of the latter.

There is growing evidence that aberrant wiring plays a central role in the etiology, pathophysiology, and symptomatology of schizophrenia [8], autism [9], and dyslexia [10]. Patients with autism and other pervasive developmental disorders are observed to have reductions in the size of the corpus callosum [11,12] and in long-range frontal/temporal functional connectivity [13,14,15]. Autism is thought to be highly heritable and polygenic [16], and a number of mouse genetic models have been developed. The ability to compare the connectivity phenotypes of different mouse models with wild-type mice could yield important clues regarding the common pathways for generating the behavioral phenotype. Currently, however, the baseline connectivity data required to make such data-driven comparisons is lacking. If connectivity phenotypes can be established for autism and other disorders, these can assist in screening for drug development and more accurate sub-typing of psychiatric diagnoses.

The importance of circuit considerations for differentially characterizing disorders such as major depression, anxiety and obsessive-compulsive disorders, and substance (including nicotine) addiction is beginning to be recognized. These illnesses are considered disorders of the affective circuitry underlying emotion and motivated behaviors, which spans the brainstem, hypothalamus, frontal and cingulate cortices and basal cortical nuclei [17,18]. Knowledge of affective circuits is substantially poorer than of sensory-motor circuitry, despite disorders of the former resulting in a much greater burden of illness. Determining connectivity in these systems will allow the development of objective diagnostic tools, and may also yield cross-mammalian emotional endophenotypes to guide new conceptualizations of core psychiatric syndromes and aid drug discovery [19]. The development of animal models that mimic neuropsychiatric disorders at the circuit rather than behavioral level may also facilitate new therapeutic strategies. Furthermore, neuropsychiatric disorders likely result from pathologies at the system level, with complex genetic, epigenetic and environmental factors combining to impact the neural circuitry. Systems-level knowledge, including neuroanatomic connectivity, may thus prove crucial in better understanding results from, for example, genome-wide association studies. Analogously, the importance of incorporating knowledge from cellular systems biology (e.g. by grouping genes into pathways) has been recognized in other domains.

*What is being proposed?*

We propose a concerted experimental effort to comprehensively determine brainwide mesoscale neuronal connectivity in model organisms. Our proposal is to employ existing neuroanatomical methods, including tracer injections and viral gene transfer, which have been sufficiently well established and are appropriately scalable for deployment at this level. The first and primary objective is to apply these methods in a standardized, high-throughput experimental program to fully map the mesoscale wiring diagram for the mouse brain and, following the model of successful genome projects, to rapidly make the results and digitized primary data publicly accessible. The second objective is to collate and, where possible, digitize existing experimental data from the macaque, and to pursue targeted experiments using standardized protocols to plug key gaps in knowledge of primate brain connectivity. Additionally, we argue for similar efforts in other model organisms and for the pursuit of experimental methods that can be used in post-mortem human brain tissue.

The projects may be carried out in a distributed manner by coordinating efforts at multiple experimental laboratories making use of uniform experimental protocols, or in a more centralized way by creating one or a few dedicated sites. Here we outline the properties of a large-scale connectivity mapping project that are seen as essential, and some that are desirable but not required. The required attributes are as follows:

1. *Brainwide coverage at a mesoscopic resolution*: The experimental technique must be applicable in all brain systems, cortical and subcortical. It should not be applicable only to specific cell types; if the technique is used to target specific cells, it must be capable of targeting any cell group.
2. *Validated and extensible experimental techniques*: The experimental methods must be well characterized and, to the extent possible, validated. The *false positive* rate should be especially low. The techniques must be amenable to high-throughput application; the individual steps for sample preparation, injection, histology, detection and data analysis should be stereotyped and of limited complexity.
3. *Centralized, open access data repository*: The data collected from such an effort must be made freely available to all researchers from a centralized data repository. This includes raw image data, processed summary data and metadata.
4. *Compatibility with existing neuroanatomical resources*: The results of this project must be interpretable with respect to existing data sets. For example, creating ties to the ABA [3], existing connectivity databases (Table 1), and other atlas projects [e.g. 20,21,22] is imperative.
5. *Tractability with current informatics technology*: The data collected and maintained in the repository must be suitable to be analyzed and stored using existing informatics techniques and available technology, allowing for predictable growth in both methods and hardware.

Additional characteristics that would enhance the project's impact include:

1. *Availability of detailed anatomical information*: The ability to characterize various additional properties of the observed projection patterns would be beneficial. This might include classification of the neuronal cell types and neurotransmitters involved, laminar origins and terminations of projections in stratified structures, receptor information, cell density estimates in the origin and termination areas, morphological properties of the axons and/or dendritic arbors, and statistical characterization of topography and convergence or divergence patterns of projections.
2. *Reconstruction of projection trajectories*: In addition to the origins and terminations of projections, it would be valuable to determine their spatial trajectories. Such data would be particularly useful, for example, in understanding the impact of white matter lesions.
3. *Compatibility with high-resolution methods for targeted investigations*: While the primary imaging data should be obtained with light microscopy, electron microscopy or other high-resolution imaging methods could enable more detailed study of particular systems, provided the experimental protocols remain compatible with such techniques.
4. *Characterization of inter-subject variability*: As discussed above, quantifying the variability of observed connectivity patterns would be valuable. This would require additional informatics methods and a substantially larger number of experiments than needed for estimating a single "map."

*Where are we now?*
Assessing the extent of current connectivity knowledge in various species is difficult because virtually all aspects of previous reports, including the specifics of animals used, experimental methodology, anatomical nomenclature, and presentation of results have varied across studies and laboratories. Furthermore, published data often include only processed results in the form of prose, tables, and schematic illustrations while primary materials including original tissue sections sit on laboratory shelves.

A small number of public repositories for connectivity information are available (see Table 1), including two major efforts to manually curate reports for specific species. The CoCoMac database catalogs axonal tracing studies from the monkey literature [23,24] (approximately 400 literature reports detailing ~2800 tracer injections), while the Brain Architecture Management System (BAMS) focuses on the rat [25,26] (328 references describing about 43,000 reported connections). Both systems organize connections based on discrete brain regions identified by the original authors according to a particular map or anatomical parcellation and use inference engines [2,27,28] to attempt to reconcile results across different parcellation schemes and nomenclature systems. These reconciliation processes possess considerable uncertainties, and the data remain very sparse; thus any comprehensive picture of brain connectivity is not currently possible from such resources. The FACCS database [29] is a strong effort to map connectivity *data* into a common spatial framework, but is currently limited in scope to the rat cerebrocerebellar system. Our understanding of the overall architecture of model nervous systems is currently limited to very simple organisms such as the nematode *C. Elegans* [30].

Much of our theoretical knowledge of *human* brain connectivity comes from either very old sources [31] or from inference from varied reports in other species. Bürgel et al. [32] have developed a probabilistic atlas localizing major fiber bundles based on myelin staining in postmortem human brains, but these maps are very coarse and lack specificity in terms of termination zones. New technological developments such as diffusion-weighted MRI and computational techniques based on correlations in measured time series provide non-invasive methods for inferring some aspects of brain connectivity, but these methods necessarily require validation and should be complemented with more direct measurements. While an experimental program for the precise mapping of connectivity patterns in the *human* nervous system will require additional technological development, we are well-positioned to push forward with a systematic high-throughput experimental program for model organisms using mostly existing methods.

*A survey of available techniques*
Reviews of the history [33] and application of various techniques for determining anatomical connectivity can be found elsewhere [34,35], and a further survey is presented in Supporting Text 1. Here we elaborate on methodologies suitable for the proposed experimental program.

*Neuronal tracers* allow injected molecules to be distributed within intact living neurons through active intra-axonal transport mechanisms. Tracer substances (see Supporting Text 1 for further details and properties) can be described by their preferred direction of transport, although labeling is often not exclusively unidirectional. Importantly, the majority of neuronal tracers can only be transported *within a cell* and do not cross the synapse; their utility in revealing the connectivity between brain areas is in tracing projection neurons either from axon terminals to potentially distant cell bodies or vice versa. Retrograde transport (from axon terminal to cell body) is used to label the cells projecting to a particular target region, while anterograde transport (from cell bodies to axon terminals) allows for labeling the projection terminal regions of a cell or group of cells.

Modern "conventional" tracers yield strong, high-resolution labeling of fine processes, and can often be used in combination with one another, with histochemical techniques, genetic markers, light or electron microscopy, and a variety of delivery mechanisms. While there are many tracers that may prove suitable in a large-scale connectivity mapping project, *phaseolus vulgaris* – leucoagglutinin (PHA-L) [36] and high molecular weight (10 kDa) biotinylated dextran amines (BDA) [37,38], both of which have now been used extensively and are transported primarily in the anterograde direction over sufficiently long distances, are strong candidates for high-throughput use. Either tracer can also be used in conjunction with a second high-resolution tracer such as cholera toxin subunit B (CTB) [39], which is transported primarily in the retrograde direction, in a multiple labeling protocol [40,41]. Such multi-tracer methods allow a single experiment to be used to probe the inputs and outputs for a particular injection site at a relatively low additional cost in the detection process.

Some tracer substances, and in particular *neurotropic viruses* such as rabies virus [42] and the alpha herpes viruses [43,44], can be transported trans-neuronally to label either pre- or post-synaptic cells. Viruses enter first-order neurons, replicate, and are transferred at or near the synapse to second-order cells where replication occurs again, thus continuing a self-amplication process. Viral spread, however, has a variable time course (which depends on projection strength), thus often making, for example, differentiation of weak first-order and strong second-order projections difficult, although this problem may be alleviated by using genetically engineered viruses that cross only a single synapse [45].

Replication incompetent *viral vectors* engineered from adeno-associated virus (AAV), lentivirus, herpesvirus and others can be used to drive high expression of fluorescent proteins as anterograde and retrograde tracers. These methods can have higher sensitivity than conventional tracer methods [46,47,48]. In addition, the number and types of infected neurons can be characterized, facilitating the pooling of data across multiple experiments. These viral reagents can be used in combination with transgenic mouse lines to label specific cell types [49,50,51]. It is clear that these and other genetic techniques will continue to gain prominence in neuroanatomy [52].

*How will we get there?*
*Mouse*: The first and primary phase of our proposal is to systematically map mesoscale connectivity in the mouse brain using standardized methods to label neuronal projections in combination with optical microscopy. The mouse, as opposed to rat, is the preferred rodent model due to its increasing use in neuroscience [53], the ease of use of transgenic methods, and the availability of large-scale spatial gene expression data in the brain [3,22]. Accordingly results from the mouse will be readily reconcilable with existing data, and new anatomical methods should be quickly applicable, supplying diverse information to supplement the initial experiments. A sample workflow, timeline, and cost estimates for a comprehensive mouse connectivity project are included in Supporting Text 2. We estimate that the complete mouse project can be completed in 5 years at a total cost of less than 20 million dollars, using five replicated experimental pipelines, each consisting of uniform experimental equipment with technicians implementing standardized protocols. Increasing the number of pipelines would proportionately reduce the timeline.

The proposed protocol calls for systematic injections of conventional tracers and/or viral vectors in the young adult mouse, age- and weight-matched to an existing stereotaxic brain atlas. The ABA has established a standard by using male 56-day old C57BL/6J mice [3], and this group has developed a corresponding anatomical reference atlas that is a reasonable choice to be adopted for this project. It is vital that the mouse strains, ages and atlases used are common across the project. Furthermore all surgical procedures, injection methods, histological techniques and experimental apparatus should be uniform to reduce variability in results. The use of motorized stereotaxic manipulators with encoded positions relative to standard landmarks, and the incorporation of automation where possible within the experimental protocols will greatly aid this task. Equipment is now available for automated or semi-automated scanning and digitization of labeled sections at sub-micron resolution using fluorescence or bright field microscopy (see Supporting Text 1) and will form a critical piece of the experimental pipeline. Digitized images will be transferred into a distributed data processing pipeline for automated analysis of the experimental results and entry into a public database.

The project will necessitate further development of algorithms to reliably extract wiring information from digitized images, and to bring data from different sections and animals into register with one another. Photomicrographs from an individual animal must be registered in 3D while accounting for tissue distortions, a process that can be improved by acquiring low-resolution reference block face images prior to cutting each section [54]. Detection of labeled cell bodies or clusters of cells and 3D registration to a Nissl-based atlas are problems that have been previously addressed at a large scale, for example, in the ABA [55]. Detection of labeled axonal segments is somewhat more challenging, and typically relies on (sometimes software-assisted) manual tracing, but progress has been made towards providing automated, quantitative estimates of axon length and density [56,57]. Importantly, the objective of the analysis stream need not be to reconstruct individual neurons, but rather to detect and quantify labeled areas outside of the injection site and represent those data in a common framework [see also 58].

*Primate*: A high-throughput investigation in primates, on the scale proposed for mouse, is not feasible. Primate experiments are tremendously more costly, and the monkey brain is considerably larger, more complex, and more variable than mouse. It is therefore of critical importance that: 1) results from previous connectivity studies in primates are carefully curated from the existing literature, leveraging ongoing efforts such as CoCoMac [23,24], 2) efforts are made to systematically digitize slides that remain available from previous studies following such efforts as BrainMaps.org [59], and 3) targeted experiments using standardized protocols are put in place that yield maximal data to validate and "fill in" the mesoscopic connectivity matrix for the Macaque. See Supporting Text 3 for further details for a proposed primate connectivity project.

*Informatics considerations*: The success of the proposed efforts will hinge on the ability to coordinate activities across laboratories while maintaining quality control, to automate the analysis of acquired data, to store both raw and processed data, and to make the integrated results reliably available to different user groups through intuitive interfaces. Management of the large-scale data set will require significant computational equipment and informatics expertise, some of which is likely to be distributed across multiple sites. The scope of the proposed project demands a customized laboratory information management system (LIMS) to organize and track tasks and materials within and across sites. Much can be learned from the informatics procedures carried out at the Allen Institute for Brain Science [55] and from the significant data sharing efforts in genomics and bioinformatics [60,61].

A major challenge is to develop an appropriate structured database to store the results of injection experiments, digitized legacy data, and associated metadata. In the CoCoMac and BAMS databases, the underlying data model of anatomy is discrete; that is, each "connection" is associated with a pair of discrete brain sites. Through systematic injections, and by preserving and storing primary image data, it is possible for the underlying data to be represented in analog form. Spatial databases [62] as used in geographical information systems and, in some cases, neuroscience [63] provide many of the necessary tools once the underlying data model (e.g. coordinate system) has been established. Anatomical parcellations based on different atlases may then be probabilistically registered to this coordinate space to enable the representation of the full connectivity data in the form of connectivity graphs or matrices, with "nodes" defined by the particular parcellation. The SumsDB database (http://sumsdb.wustl.edu/sums), for example, includes a surface-based macaque atlas containing many anatomical partitioning schemes registered to a common spatial framework, along with maps of neuronal connectivity from retrograde tracer injections registered from individual subjects to the atlas [64]. Representation in the continuous space additionally allows for a post-hoc analysis that *solves for* the partitioning of brain space that best follows the connectivity patterns observed in the data.

*Technology development and evaluation for human studies:* The ultimate goal of our proposal to experimentally map brainwide connectivity patterns is to arrive at a comprehensive understanding of the architecture of the *human* brain. A much improved partial understanding can be obtained from the proposed efforts in mouse and macaque, and a proposal has been made for a human connectivity project that would rely primarily on neuroimaging techniques [65]. Still, resources should be devoted to developing classical neuroanatomical techniques that are viable for humans. There have been sporadic efforts to increase the speed of action for lipophilic carbocyanine dyes when used in fixed postmortem human tissue [66,67], and these and other efforts should be studied further. Additionally, imaging methods including diffusion tensor imaging and diffusion spectrum imaging, as well as computational techniques for the assessment of "functional" or "effective" connectivity [68] can be validated by supplementing tracer studies in macaque with MR data acquisition in the same animals. Such efforts are essential to ultimately reversing the backwardness of human neuroanatomy.

## *Conclusions*

The largest current gap that exists between the genotype and phenotype in neuroscience is at the level of brain connectivity. There is thus enormous potential value in the acquisition of comprehensive, unified connectivity maps in model organisms. We have proposed a concerted effort within the neuroscience community to determine these connectivity patterns *brainwide* at the tractable yet representative mesoscopic scale, first in the mouse, and followed by additional efforts for the macaque and eventually humans. The mouse proposal is based on existing methods, scaled up, and standardized for high-throughput experimentation. This effort would be complementary to, and would provide "scaffolding" for additional anatomical projects using different emerging technologies, and can be integrated with existing resources such as the ABA to probe various levels of structural and functional organization. Examination of a potential project plan demonstrates that such an effort would be relatively inexpensive in terms of both money and time (see Supporting Text 2) compared to its potential value in neuroscience and biomedicine. If successful, similar projects could be undertaken for model organisms including vertebrates such as rat, zebrafish, zebra finch and chick and invertebrates such as jellyfish, flatworms and drosophila, enabling comparative neuroanatomical studies that are currently well beyond reach.

While the principal objective of the proposed project is to characterize and make available a "wiring diagram," the public availability of raw data is vital to allow researchers to form their own, perhaps more detailed, interpretation of the individual results. Technological advances have only recently made it feasible to capture and store the voluminous raw image data at sub-micron resolution, and to serve these images over the web. The spirit of collaboration and open data access requisite in this proposal is also currently reflected in increasing proportions within the neuroscience community and within funding agencies as reflected, for example, in the NIH Blueprint for Neuroscience Research [69] and in international neuroinformatics initiatives [70]. Thus we may be at a point in time that makes a project of this sort uniquely feasible. Realizing the vision put forth here will require additional planning, input from the community, and financial support. Moreover, eventually determining the connectivity matrix for *human* will require additional technical development. The hope, however, is that this proposal has made both the importance and viability of brainwide connectivity projects apparent, and that we can move from planning to action on a short time scale.

*Acknowledgments*

This paper was prepared as part of the Brain Architecture Project (funded by the W.M. Keck Foundation) at Cold Spring Harbor Laboratory. P. Mitra. L. Swanson, J. Doyle, H. Breiter and C. Allen were instrumental in the planning and assembly of the Brain Architecture Project. The paper is a result of discussions at the 2007 and 2008 Banbury Center meetings with a working group of scientists who provided input and editions to the manuscript, and who are listed alphabetically as middle authors. The manuscript was conceived by P. Mitra and written by P. Mitra and J. Bohland with much assistance from C. Wu. Significant contributions to the paper were also made by L. Swanson, H. Barbas, S. Haber, C. Hilgetag, R. Kötter, and K. Svoboda.

*Figures*

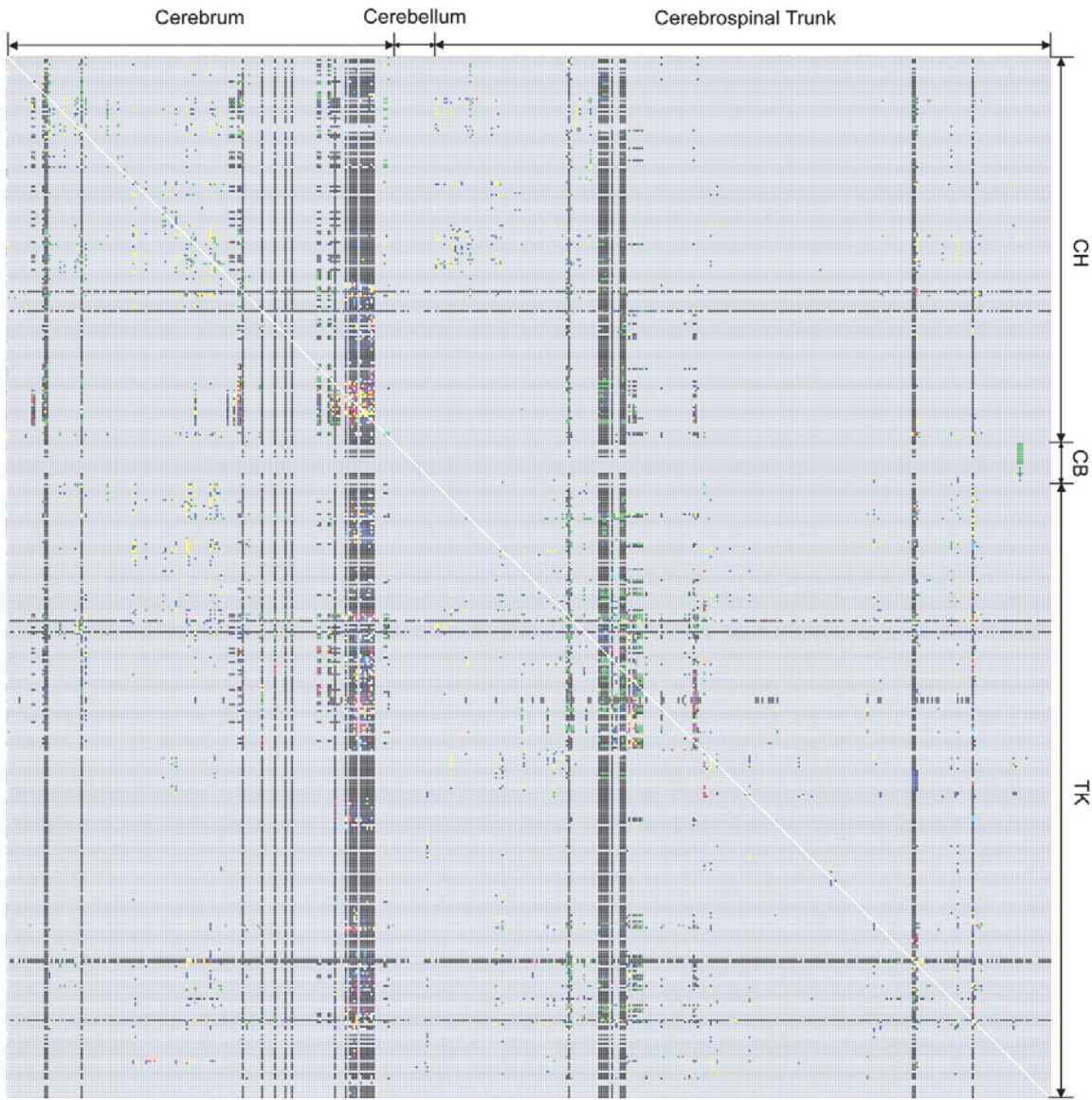

**Figure 1:** Knowledge of rat brain connectivity available in the Brain Architecture Management System. This matrix shows information that has thus far been curated about projections between 486 discrete brain regions in the rat brain. Reprinted with permission from [Ref. 2]. Non-gray entries indicate connections for which data are available. Black entries indicate the absence of a connection, and colored entries indicate reported connections of varying strength. The overall sparsity of this matrix is reflective of our lack of a unified understanding of brain connectivity in model organisms.

**Table 1.** Databases and datasets containing information about neuroanatomical connections.

| Database | Available connectivity information | URL |
|---|---|---|
| Brain Architecture Management System (BAMS) [2,26] | Projections in rodent brain, curated manually from existing literature | http://brancusi.usc.edu/bkms/ |
| Collations of Connectivity data on the Macaque brain (CoCoMac) [23,24] | Projections in Macaque brain, curated manually from existing literature | http://www.cocomac.org |
| Functional Anatomy of the Cerebro-Cerebellar System (FACCS) [29] | 3D atlas of axonal tracing data in rat cerebro-cerebellar system | http://ocelot.uio.no/nesys/ |
| BrainMaps.org [59] | Tables of connections from literature and primary data for some tracer injections | http://brainmaps.org |
| BrainPathways.org | Multi-scale visualization of connectivity data from collated literature reports | http://brainpathways.org |
| Human Brain Connectivity Database | Curated reports of connectivity studies in post-mortem human brain tissue | http://brainarchitecture.org |
| Internet Brain Connectivity Database | Estimated connectional data between human cortical gyral areas | http://www.cma.mgh.harvard.edu/ibcd/ |
| SynapseWeb | Reconstructed volumes and structures from serial section electron microscopy | http://synapses.clm.utexas.edu/ |
| Neocortical Microcircuit Database[71] | Connection data between single cells in mammalian cortex | http://microcircuit.epfl.ch/ |
| ICBM DTI-81 Atlas [72] | Probabilistic atlas of human white matter tracts based on diffusion tensor imaging | http://www.loni.ucla.edu/Atlases/Atlas_Detail.jsp?atlas_id=15 |
| Anatomy Toolbox Fiber Tracts [32] | Probabilistic atlas of human white matter tracts based on postmortem studies | http://www.fz-juelich.de/ime/spm_anatomy_toolbox |
| WormAtlas [30] | Full neuronal wiring data for *C. Elegans* | http://www.wormatlas.org |

**A proposal for a coordinated effort for the determination of brainwide neuroanatomical connectivity in model organisms at a mesoscopic scale: Supporting information**

## Supporting Text 1: Survey of methods relevant for determining neuronal connectivity

To supplement the discussion provided in the main article, here we provide a brief general overview of experimental methods for determining and imaging neuronal connection patterns.

*Classical methods*: The most basic class of anatomical methods aimed at connectivity, which are based on gross dissection, have been used roughly since the initial distinction between white and grey matter was made by Vesalius [1], with the "teasing method" for following the course of fiber bundles developed by Steno in the 17$^{th}$ century effectively inaugurating the era of tract tracing. Gross dissection provided a vehicle for an elementary understanding of white matter structures, but is severely limited by an overall lack of precision in defining anything but the largest pathways and an inability to map the fibers to their precise terminations.

Innovations in staining methods [2] as well as the development of neuronal degeneration techniques [3,4] led to Dejerine's comprehensive study of the human white matter pathways, still frequently referenced today [5]. In later years such methods were improved to selectively stain degenerating fibers [6,7], and these techniques have provided much of what is known even today about the fiber systems in humans. Still, degeneration methods are severely limited due to a lack of selectivity, and because they require insult to the neural tissue under investigation. Furthermore these methods are tedious and labor-intensive and are not feasible for any large-scale attempt to determine brainwide connectivity patterns at an appropriate level of resolution.

*Diffusion and functional imaging:* Non-invasive brain imaging techniques are now the most common tools for modern *human* neuroanatomy. Diffusion tensor imaging (DTI) [8] can be used to estimate the tensor describing water diffusion at each voxel in a magnetic resonance image (MRI) volume. This technique is of use for assessing connectivity because diffusion within the white matter is constrained by axonal membranes with particular orientations. Various tractography algorithms can be used to "connect" the tensors along their principal directions to give an estimate of the orientations of fibers traversing through the volume. Studies have demonstrated that such techniques can yield the approximate trajectories of a number of major fiber bundles that have been demonstrated via other methods in either human or non-human primates [9,10]. DTI, however, is limited in its specificity where fibers enter the gray matter, and also because of considerable ambiguity at potential crossings of fiber tracts within the volume. Diffusion spectrum imaging (DSI) [11] reduces the problem of fiber crossings by providing a much finer sampling of the orientation distribution function at each voxel.

Another class of inferential methods is based on analysis of correlations in multivariate time series data obtained from either direct measurements using, for example, multi-unit electrodes, or from indirect measurements using electro- or magnetoencephalography (EEG/MEG) or functional MRI. A wide variety of statistical techniques are now available to infer functional networks from these correlations [12,13,14,15], yet the inference of *anatomical* connectivity from task-based functional data is tenuous. Indeed the availability of mesoscopic connectivity information would greatly constrain efforts to measure task-related modulations of functional or effective connectivity in imaging studies. There has been recent interest in spatially correlated activity observed in *resting state fMRI* [16,17,18] where the subject is not actively performing a task. Results suggest that spontaneous correlated spatial patterns may reflect anatomical subdivisions [19], but further data are needed to understand the complex relationship between correlations in the blood oxygenation level dependent (BOLD) response and physical connections between the neurons indirectly giving rise to that response. Still, brain imaging methods provide truly non-invasive techniques for the study of connectivity, especially in humans, and may prove quite valuable, particularly for longitudinal studies. An important contribution of a large-scale connectivity project using other more well-established methods in non-humans can be to help validate these methods (see the primate project described in Supporting Text 3).

*Neuronal tracers:* A great advancement in neuroanatomy occurred as tracers based on axonal transport were rapidly introduced beginning in the 1970s, allowing injected molecules to be distributed within intact living neurons through active intra-axonal transport mechanisms. In their typical use, following a variable post-injection survival period, the animal is sacrificed, the brain tissue processed, and tracer detected through histochemical reactions and microscopy. The versatility of neuronal tracers has made them the primary method for examining anatomical connectivity over the past several decades. The first tracers used radiolabeled amino acids [20] incorporated into cell proteins and transported in the anterograde direction. The transported radiolabeled isotopes can then be detected using autoradiography. This method has been used extensively in monkey studies, but suffers from an indirect labeling method and a lack of resolution compared with more recent tracers. Retrograde transport was achieved by injection of the enzyme horseradish peroxidase (HRP) [21,22], which could be robustly detected with a simple histochemical reaction using, for example, diaminobenzidine (DAB) or tetramethylbenzidine (TMB) as a substrate.

More recently many improved "conventional" tracer substances that result in stronger or higher-resolution labeling have been developed (see Table 1). We discuss two such tracer substances with preference for anterograde transport in the main article and in more detail presently. *Phaseolus vulgaris* – leucoagglutinin (PHA-L) [23] is a lectin derived from red kidney beans that binds to glycoproteins in neuronal membranes and is taken up into cell bodies and is transported almost exclusively in the anterograde direction. Biotinylated dextran amines (BDA) [24,25] are dextrans, which have been used in other forms as tracers, that are conjugated to biotin. Available in different molecular weights, the heavier versions (e.g. BDA10k; 10 kDa MW) appear to show stronger preference for anterograde transport, whereas retrograde transport is observed for lighter versions [26]. Still BDA is not transported in either direction exclusively, which can lead to some ambiguity in interpretation of labeling results; this problem is minimized with PHA-L. The incorporation of biotin in BDA enables a simple histochemical detection procedure that is suitable for light or electron microscopy using a standard avidin-biotinylated HRP (ABC) kit followed by DAB reaction, whereas the detection of PHA-L requires incubation in primary and secondary antibodies. Both can be injected by iontophoresis, preferred for small injection sites, or by pressure injection for larger targets.

We have suggested that the use of a multi-tracer protocol [27,28] involving co-injection of an anterograde and retrograde tracer may prove beneficial in a high-throughput experimental program. The non-toxic B subunit of cholera bacterial toxin (CTB) [29] is a strong candidate for use as such a (primarily) retrograde tracer. Alternatively, the fluorescent tracer Fluoro-gold [30] is also suitable for combination with other tracers in rodents [31]. Multiple labeling protocols require that each tracer be revealed as a distinct reaction product; this can be easily achieved with an appropriately designed series of standardized histochemical processes.

It is outside of our purposes to discuss the merits of each neuronal tracer (though see Table 1), but another class – the lipophilic carbocyanine dyes (including DiI, DiO, and DiA) – provide at least one property not available to the majority of other methods: the ability to act in fixed post-mortem human brain tissue through passive diffusion [e.g. 32]. While the slow diffusion process has thus far limited their use to the study of relatively short-range (up to 2-3 cm) projections, attempts have been made to increase the speed of transport to ultimately make the method more viable for human tissue [33,34]. These and other fluorescent tracers can be injected *in vivo* as well, but have few benefits relative to the other tracers discussed above, and suffer from fading of the fluorescence signal after exposure to light. The labeling can be made into a permanent substrate through photoconversion, but the process is time consuming and not viable for high-throughput application.

*Trans-neuronal tracers*: Some tracer substances can be transported trans-neuronally by crossing the synaptic cleft to label either pre- or post-synaptic cells. Among "conventional" tracers these include the non-toxic C fragment of tetanus (TTC), wheat germ agglutinin (WGA) and, in some systems, WGA conjugated to horseradish peroxidase (WGA-HRP). These tracers become increasingly dilute as they spread, resulting in the problem of weak labeling and difficult detection. The use of neurotropic viruses has largely superseded the utilization of such tracers for labeling multi-synaptic pathways (see Table 2). The two major classes of viruses in use as trans-neuronal tracers are rabies virus [35] and the alpha herpes viruses [36,37] including the swine pseudorabies virus (PRV; not related to rabies virus) and herpes simplex virus (HSV-1). Viruses enter the cell bodies of first-order neurons, replicate, and are then transferred at or near the synapse to second-order cells where replication occurs again. The virus thus has the important property of self-amplification, which results in generally superior labeling of neurons compared to the non-viral trans-neuronal tracers. Because virus spread across multiple synapses has a variable time course (which is affected by the strength of projections, for example), it is often difficult to differentiate weak first order connections from strong second-order projections, and so on. For this reason it is typically necessary to evaluate viral labeling (using immunohistochemical procedures) at multiple time points following injection, requiring additional injections and additional animals.

*Transgenesis and viral gene delivery*: One of the primary criticisms of conventional tracer techniques is their general inability to target specific cell types. The incorporation of genetic methods into the neuroanatomist's toolbox [reviewed in Refs. 38,39] has begun to bridge this gap as well as to augment the conventional techniques. Here we give only a few illustrations of these techniques. For example, WGA cDNA has been employed as a transgene in mice, with expression under the control of cell-type specific promoters [40,41] in order to label selective multi-synaptic circuits in the anterograde direction. TTC fused with green fluorescent protein (GFP) expression has similarly been engineered in transgenic mice for retrograde tracing [42]. These genetically targeted trans-neuronal tracers still, however, suffer from the same shortcomings of their conventional counterparts, particularly low sensitivity to weak connections.

Viral tracers have additionally benefited from genetic manipulation. The PRV virus was engineered to be cell-type specific by replacing the coding sequence for thymidine kinase (TK), a gene necessary for replication, with a conditionally expressed TK gene as well as the coding sequence for GFP [43]. The modified virus then is only able to replicate in cells expressing Cre recombinase, and these cells also express the fluorescent protein. When the virus is recombined, it spreads as usual in the retrograde direction, with infected cells in the multi-synaptic pathway also expressing GFP. Thus this method allows the initial first-order target of the virus to be restricted to cell populations that themselves can be targeted for Cre expression through transgenic methods. A recent innovation was made by Wickersham et al. [44] to restrict trans-synaptic labeling by a deletion mutant rabies virus to a single synapse, beginning from a targeted cell population. The virus envelope protein (rabies glycoprotein; RG), necessary for trans-synaptic spread but not for transcription or virus replication, was replaced by enhanced green fluorescent protein (EGFP). The missing virus glycoprotein gene is then supplied *in trans* to a cell group targeted for initial infection with the mutant tracer virus. The result is that the virus can assemble and spread trans-synaptically from this initial cell population, but because RG is not present in the pre-synaptic cells, the virus can not continue to spread. Infected cells also strongly express EGFP making detection with fluorescence microscopy straight-forward.

Replication incompetent viral vectors engineered from adeno-associated viruses (AAV), lentivirus, and others can be used to deliver genetic material to selected cells [45], notably to drive high expression of fluorescent proteins as anterograde and retrograde tracers. Recombinant AAV [46] and lentivirus [47,48] have both, for example, been used to deliver GFP to cells by injection, resulting in a robust anterograde labeling of axons over potentially long distances that is stable over time. This method thus acts much like a "conventional" anterograde tracer, with potentially higher sensitivity, and can be visualized with fluorescence microscopy or with bright-field microscopy after appropriate antibodies. Additionally lentivirus has been engineered with GFP-tagged synaptophysin to produce selective labeling at presynaptic terminals [47]. These viral vectors can additionally be used in combination with transgenic mouse lines to target specific cell types [49,50].

Recently *Brainbow* transgenic mice were developed also exploiting the Cre-Lox system [51]. Through engineered combinatorial expression of multiple fluorescent proteins, targeted neurons take on many (up to 90 in the published report) distinct color profiles. The color variations allow one to distinguish adjacent neurons and thus to limit ambiguity in the reconstruction of axons and synaptic contacts. While not a tracer method, the *Brainbow* construct offers the ability to probe detailed microcircuitry in relatively large groups of labeled neurons; furthermore automation of neuron reconstruction should be made easier by the distinct color profiles, which have been shown to be relatively stable within each cell.

*Microscopy methods*: Any of the neuronal tracing techniques available require a stage in which labeled cell bodies or neuronal processes are revealed through an imaging process. This has been typically achieved through light microscopy (LM), which current remains the most viable option for a high-throughput setting. Image capture for the Allen Institute's gene expression atlas project was fully automated using Leica DM6000B microscope systems (http://www.leica-microsystems.com/) [52], and the Brainmaps.org project [53] has used systems from Aperio Technologies (http://www.aperio.com) for "virtual microscopy." These types of automated imaging technologies, which have only become available in recent years, can offer very high (less than 1 square micron) in-plane image resolution and will be suitable for detection of reaction products resulting from immunohistochemistry. In addition to systems from Leica and Aperio, the NanoZoomer from Olympus (http://www.olympusamerica.com/), and the Mirax Scan System from Carl Zeiss, Inc. (http://www.zeiss.com) offer similar technologies.

Such LM technologies, however, are insufficient for extremely high-resolution imaging of synapses and receptors or other subcellular organelles. Recent developments in ultrastructural imaging including array tomography, serial reconstruction of pathways[54] at the site of termination [e.g. Ref. 55], serial block-face scanning electron microscopy (SBFSEM), and the automatic tape-collecting lathe ultra-microtome (ATLUM), have opened the possibility of obtaining accurate large volume images at nanometer resolution [56,57,58]. Such methods are not tractable for comprehensive brainwide mapping in large brains, and it is unclear that the required neuron reconstruction algorithms ready for high-throughput usage. Further, some elements of microcircuit organization may be inferred without ultrastructural reconstructions using principles such as Peter's Rule [59]. However, these techniques should be used to provide detailed supplemental data to the proposed project. For this reason it is important that techniques applied to a brainwide project remain compatible with EM.

| Procedure | Used in | Spatial specificity | Brief Description |
|---|---|---|---|
| Gross dissection (1543) | postmortem human | major fiber bundles | classical "hands on" anatomy |
| Myelin stains (1882) | any species | large and fine myelinated axon bundles | ferric chloride and hematoxylin stain myelinated fibers deep blue |
| Degeneration Methods (1885-1950's) | any species | degenerating axon bundles | e.g. silver impregnation selectively labels degenerating fibers from a lesion site |
| Diffusion-based MR imaging (1990 - ) | human | major fiber bundles | MR imaging of water diffusion direction + reconstruction |
| Neuronal tracers (1968 - ) | rodents, primates, postmortem human | Single neurons and/or fibers down to neural processes | active or passive transport of injected substances |
| Transgenesis (1990 - ) | mice | fiber bundles down to neural processes | selective expression of genes to label specific cells |
| Viral gene transfer (1998 - ) | any species | fiber bundles down to neural processes | selective expression of genes to labels specific cells |
| SPIM / Ultramicroscopy (2007)[†] | any species | single cell | planar imaging of whole sample |
| Volume EM reconstruction (2004 -)[†] | any species | nanometer resolution; single boutons, spines, receptors and other subcellular organelles | serial scanning of ultrathin sections, or block-face scanning of three-dimensional volumes |

(Left margin: arrow from *coarser* to *finer*)

**Table 1:** Comparison of conventional neuronal tracer substances. [†] Imaging methods (not connectivity methods *per se*) that can be used in conjunction with dyes and tracers, etc. for detailed reconstruction of microcircuitry.

| Tracer | Labeling efficacy | | Resolution | Application | Detection | Visual-ization | Mechanism | Trans-synaptic | Transport rate | Preparations | Limitations |
|---|---|---|---|---|---|---|---|---|---|---|---|
| | A | R | | | | | | | | | |
| Radiolabeled amino acids[20] | ■ | | Medium | S | Autoradiography | LM, DFM | Protein synthesis and fast axonal transport | | up to 100 mm/day | In vivo primate, rodent | Inferential detection method |
| Horseradish peroxidase (HRP)[21,22,60] | ▨ | ▨ | Medium-Fine | P, I, S | ABC method | LM, EM | Endocytotic uptake, retrograde vesicular transport | | 200-300 mm/day | In vivo rodent, primate | Tracer leakage, lack of fine labeling |
| Lipophilic Dyes (DiI, FastDiI, DiA, DiO)[61] | ▨ | ▨ | Medium-Fine | P, I, S, C, B | Red, green, blue fluorescence, photoconversion | FM, LM | Lateral diffusion within fluid membrane | | up to 6 mm/day in vivo, ~2 mm/mo in fixed tissue | In vivo rodent, primate, slice cultures, fixed tissue, birds | Photobleaching, failure in adult animals |
| Fluoro-Gold (FG)[30] | | ■ | Fine | P, I, C | gold-blue fluorescence, photoconversion, IH | FM, LM, EM | Endocytotic uptake, retrograde vesicular transport | | | In vivo rodent | Limited use for long-term studies; cytotoxicity |
| Phaseolus vulgaris-leucoagglutinin (PHA-L)[23] | ■ | | Fine | I | IH | LM, EM | Binds to cell surface receptors | | 4-7 mm/day | In vivo rodent | Not effective in older animals, unreliable for some researchers in primates |
| Wheat germ agglutinin (WGA)[62] | ▨ | ▨ | Medium | P, I, T | IH | LM, EM | Binds to cell surface receptors | Y | 22-44 mm/day | In vivo rodent, primate, transgenic mice | Severe immune response, weak transneuronal labeling |
| Cholera toxin subunit B (CTB)[63] | ▨ | ■ | Fine | P, I | IH | LM, EM | Binds to cell surface receptors | | ~100 mm/day | In vivo rodent, primate, birds | Some anterograde transport |
| C fragment of tetanus toxin (TTC)[64] | | ▨ | Medium | P, I | IH | LM, EM | | Y | ~75 mm/day | In vivo rodent, primate | weak transneuronal labeling |
| Biotinylated dextran amines[24,25] (BDA 10kDa) | ■ | ▨ | Fine | P, I | ABC method | LM, EM | Probably endocytotic uptake followed by diffusion | | 2-6 mm/day | In vivo rodent, primate | Some retrograde transport through fibers of passage |
| Biotinylated dextran amines[24,25] (BDA 3kDa) | ▨ | ▨ | Fine | P, I | ABC method | LM, EM | Probably endocytotic uptake followed by diffusion | | 2-6 mm/day | In vivo rodent, primate | Some anterograde transport |
| Fluorescent dextrans (FR, FE, CB, LY)[65,66] | ▨ | ▨ | Fine | P, S | Multi-color Fluorescence, photoconversion, IH | FM, LM, EM | Probably endocytotic uptake followed by diffusion | | | In vivo rodent, primate | Photobleaching of label |
| Biocytin / Neurobiotin[67] | ■ | | Fine | IC, P, I | ABC method | LM, EM | Fast axonal transport | | 36-72 mm/day | In vivo rodent | Only useful for short survival periods |
| Fluorescent latex microspheres[68] | | ▨ | Medium | P, S | Multi-color Fluorescence | FM, LM | Retrograde vesicular transport | | | In vivo rodent, slice culture | No cell morphology visible, limited compatibility with histological methods |

**Table 2:** Comparison of conventional neuronal tracer substances. In columns titled *Labeling efficacy*, subcolumn A indicates anterograde direction, R retrograde direction, and the gray-level indicates efficacy with black strongest. In column *Application,* S: Hamilton microsyringe, P: pressure injection, I: iontophoresis, C: crystal placement, B: biolistic delivery, T: transgenic mice. Other acronyms: IH: immunohistochemistry, LM: light microscopy, FM: fluorescence microscopy, EM: electron microscopy, DFM: dark field microscopy, ABC method: avidin biotinylated HRP method.

| Virus | Direction A | Direction R | Specific targets | Trans-synaptic | Reporter Method | Genetic constructs required | Human pathogen | Known issues / considerations |
|---|---|---|---|---|---|---|---|---|
| *alpha-herpesviruses* | | | | | | | | |
| *swine pseudorabiesvirus (PRV)* | | | | | | | | PRV viruses fail to work in primates, virus may not exclusively spread to synaptically connected cells |
| PRV-Becker | ▓ | ▓ | | multiple | IH | N | N | highly virulent wild type strain, not viable as tracer due to short survival times |
| PRV-Kaplan | ▓ | ▓ | | multiple | IH | N | N | |
| PRV-Bartha | | ■ | | multiple | IH | N | N | immune reactivity, polysynaptic, distribution of receptors, ventricle uptake |
| PRV152[69] | | ■ | | multiple | EGFP | N | N | |
| PRV614[70] | | ■ | | multiple | RFP | N | N | |
| PRV-BaBlue[71] | | ■ | | multiple | β-Gal | N | N | |
| Ba2001[43] | | ■ | Cre expressing cells | multiple | EGFP | Y | N | |
| *human herpes simplex virus (HSV)* | | | | | | | | Herpes viruses can infect astrocytes as well as neurons |
| HSV-1 (MacIntyre) | ▓ | ■ | | multiple | IH | N | Y | Transport direction may differ by infected area |
| HSV-1 (H129)[72] | ■ | | | multiple | IH | N | Y | |
| HSV-2 (Strain 186) | ▓ | ▓ | | multiple | IH | N | Y | Transport direction may differ by infected area |
| *rhabdoviruses* | | | | | | | | |
| Challenge Virus Standard (CVS-11)[35] | | ■ | | multiple | IH | N | Y | May not work in all systems in all species |
| Deletion-mutant rabies virus[44] | | ■ | TVA expressing cells | single | GFP, RFP | Y | Y | Not currently available *in vivo* |
| Recombinant Adeno-associated virus (rAAV)[46] | ■ | | | none | IF | N | N | |
| Recombinant Lentivirus[47] | ■ | | | none | GFP, IF | N | N | |

**Table 3:** Summary of a subset of the viral tracers (including recombinant viruses) that have been utilized.

A proposal for a coordinated effort for the determination of brainwide neuroanatomical
connectivity in model organisms at a mesoscopic scale: Supporting information

## Supporting Text 2: Example workflow, informatics requirements, timeline, and cost estimates for mouse connectivity project

The chief proposal that we discuss in the main article is to systematically inject conventional neuronal tracers and/or viral vectors delivering fluorescent proteins into targeted brain regions in living, young adult mice, age and weight-matched to an existing stereotaxic atlas. Here we describe in greater detail a possible experimental workflow and pipeline using conventional tracers, the details of which are only meant to be illustrative (see Figure 1). A similar workflow using the same or very similar experimental apparatus may be developed for injected viral vectors, with appropriate modifications to the various processing stages. We have called for designed redundancy in the experimental program, resulting in multiple data points per injection region; such redundancy can also, for example, allow the use of both conventional and virus-based tracers at each brain site. The full details of the experimental protocols for such a project will need to be determined through pilot experiments and further consultations with experts.

### Experimental procedures

*Protocol for injections of neuronal tracers:* Mice will be anesthetized and surgically fixed to a stereotaxic table, and a tracer injection guided by motorized micromanipulator to position the micropipette at a pre-determined location. Injection coordinates will be determined *a priori* for the full experimental program in order to approximately sample the entire gray matter. This process will require computational efforts before the experimental program begins to determine the optimal distribution of injection sites, factoring in known anatomy. A high-resolution tracer with anterograde preference (BDA in this example, or PHA-L with slight modifications) and tracer with retrograde preference (CTB) will be co-injected from the micropipette (10-20 μm diameter) using iontophoresis, roughly according to previous protocols [e.g. Ref. 1], or using microsyringes (e.g. Hamilton) and pressure injection. The injection process must be carefully controlled (electromechanically if feasible) to ensure regular injection volumes and locations, and injections will be analyzed *post hoc* to remove unsuccessful experiments. Injected mice will rest in a cage during a survival period of approximately 7-10 days to allow for sufficient tracer transport; standardized values for this and other experimental parameters can be determined during pilot studies. The animal will then be deeply anesthetized and perfused transcardially using a 4% paraformaldehyde solution, and the brain removed. The brain will be rapidly frozen and sectioned using a cryostat to cut coronal sections of ~30 μm thickness. Section plane orientation must be standardized and reproducible across experiments and pipelines, thus necessitating the use of customized brain blockers or other apparatus to assure reproducibility. Further, during the sectioning process, block face digital photographs of each section should be acquired prior to cutting [e.g. 2,3], providing an aligned reference for automated serial reconstruction algorithms (see below).

*Tissue processing:* The approximately 500 resulting sections must then be assembled into matched series of sections and mounted onto slides. A subset of sections should be stained for cellular and/or chemical microarchitecture, with the remaining slides processed for detection of the tracer substances. Sections stained to reveal cell bodies with thionin or NeuN antibody (for neuron-specific labeling) can be used *post hoc* to help localize tracer injection sites and labeled cellular processes (from adjacent sections) relative to the cytoarchitecture. These images will also have possible utility in enabling high-resolution registration of individual brains to a common template space. Additional sections can be stained for myelin using Luxol fast blue or hematoxylin or processed for other attributes such as types of neurotransmitters. Because the primary objective of the project must remain connectivity tracing, it is advisable that a large fraction of available sections (e.g. 1 out of 2-5) be used for tracer label detection, and that the remaining slides be rationed among other processes deemed valuable to the project. The number of additional histological procedures should thus be restricted to ensure that data from different modalities are within sufficient proximity to preserve the resolution of the overall data set.

*Histochemical processing:* Detection of two unique labels from the anterograde and retrograde tracer in a multiple injection protocol will proceed using a standardized procedure resulting in two distinct and permanent reaction products. BDA detection, for example, only requires incubation in avidin-HRP for approximately 90 minutes, followed by several washes in phosphate buffer (PB), then followed by incubation in DAB-Ni (diaminobenzidine containing low concentrations of nickel ammonium sulfate) for 5-10 minutes, resulting in a blue-black reaction product. CTB detection requires a longer but still systematic series of incubations, first in rabbit anti-cholera toxin antiserum for 24 hours, followed by goat anti-rabbit IgG for 60 minutes, rabbit-PAP for 60 minutes, and finally standard DAB, yielding a red-brown reaction product [see Ref. 4]. The above procedures typically require much manual skill and parameters differ from laboratory to laboratory, but in the proposed large-scale effort, they will be standardized accordingly. Additionally, automation will be added to the process where technically and economically feasible to reduce human bias and error. The Allen Brain Atlas (ABA) project, for example, demonstrated the feasibility of incorporating automation into a histological pipeline for research by using commercially-available robotics for liquid handling in their *in situ* hybridization protocol [Ref. 5; see Supplemental Methods 1]. Multiple vendors supply equipment for automated administration of staining and immunohistochemistry protocols and for coverslipping. Such technologies can reduce human bias and error within and across participating laboratories.

*Microscopy:* For the imaging stage, equipment is now available for semi-automated scanning and digitization in compressed image format of the labeled sections at sub-micron resolution using fluorescence or bright field microscopy (see also discussion of experimental methods in additional supplementary materials). Depending on section thickness and performance of microscope technologies and image processing algorithms, it may be beneficial to acquire a multiple-image stack for each section, imaged at different focal depths in the z-plane, to enable better segmentation of cell bodies and axon fibers. It is important that each site involved in the project have access to the same microscopy technologies to ensure consistency in the raw image data, and unbiased performance of signal extraction algorithms across locations. For each experiment, the complete set of treated slides (covering the entire brain) will be scanned. The digitized images will then be transferred into a distributed data processing pipeline for automated analysis of the experimental results (see below), and rapidly submitted to a public data repository that will serve not only the *results* from this project but also the *raw* image data. The physical slides can be stored locally at the site performing the experiment.

## Automated data analysis

As the primary data elicited from a high-throughput tract tracing program are images, the major informatics challenge will be to set up an automated pipeline for processing the many image

**A proposal for a coordinated effort for the determination of brainwide neuroanatomical connectivity in model organisms at a mesoscopic scale: Supporting information**

sections. The primary image processing tasks are 1) automated digitization and storage of microscopy images; 2) 3D reconstruction of anatomy from serial sections; 3) registration of different test brains to a common coordinate space; 4) detection and quantification of labeled portions of the images, and quantitative assessment of the strength and/or density of labeling; and 5) characterization of normal individual variation. The first four challenges are reasonably similar to those that have been successfully met by the ABA project [6], and these efforts should serve as a guide. The 5$^{th}$ task is an important addition to the current project, and will additionally set the stage for the study of abnormal variation in *connectivity phenotypes* using the same procedures to study mouse disease models.

Using acquired block-face digital images as a reference, the full set of sections (including those stained for Nissl, etc.) for an individual brain can be reconstructed into a 3D volume [7] with a global coordinate system. A variety of image warping methods [8,9] can be employed to register the image data for each digitized section with the corresponding block face image, matching image contours or landmarks, or by, for example, maximizing mutual information between the images [10]. This stage will need to account for alterations in the tissue shape due to histological processing, and for tissue shrinkage. After reconstruction, individual brains should be registered to a common atlas coordinate space to enable common presentation of results and analysis across individual experiments. This necessitates matching of the global shape of the individual volume with the atlas volume, which may be augmented by local image deformations to approximately match individual brain regions. The resulting deformation fields can be used to bring signals extracted from the individual section data (e.g. the axon segments and cell bodies) into the common atlas space.

A critical module needed for the proposed experimental program will contain algorithms to automatically detect and quantify labeled axonal fibers and cell bodies. Currently the usual procedure for charting labeled cells uses either a camera lucida or semi-automated image-combining computerized microscopy (e.g. the NeuroLucida system [11]), which allows the experimenter to digitally record coordinate-based data in the same space relative to an image under the microscope. With the incorporation of the stage of high-quality, high-resolution imaging of all primary data, it is quite feasible to automate the detection of the injection site and extent, as well as labeled cell bodies and fibers *throughout the entire brain* using post-hoc image analysis. The ABA employed such automated image processing steps to detect cell bodies labeled through *in situ* hybridization [6], and progress has been made toward automatically detecting and measuring labeled axonal fibers and plexuses [12-14]. These challenges amount to segmentation problems in image processing, and a wide variety of methods can be brought to bear that are likely to prove reliable in the controlled scenarios we have proposed. The output of these signal detection algorithms should include a quantitative evaluation of the location and extent of the injection site as determined by densely labeled cell bodies centered on the prescribed stereotaxic coordinates, and quantitative estimates of the fiber density and number and locations of labeled cells at projection target and source sites, respectively. Sites with very low density labeling may be ignored, as these could be due to fibers of passage or transport in the non-preferred direction resulting in labeling of axon collaterals.

**A proposal for a coordinated effort for the determination of brainwide neuroanatomical connectivity in model organisms at a mesoscopic scale: Supporting information**

**Laboratory information management system and data storage**

The proposed high-volume, multi-site experimental workflow demands an integrated, computerized laboratory information management system (LIMS) that seamlessly organizes and manages the sets of discrete tasks, protocols, and materials across laboratories. This is a critical component, particularly for projects distributed across sites, which is likely to require custom development to meet project needs. Accordingly, LIMS development and implementation will be a significant project expense, included in the coarse budget estimate in additional supplementary information. To the extent possible, the LIMS will automate planning and quality control, track materials (e.g. using bar codes), record various experimental parameters, and issue work lists to technicians and pieces of equipment at the different project sites.

The system should also link all experimental metadata with the digitized image data, which will necessitate massive, redundant data storage capabilities and computational power over a fast, distributed network. An estimate of raw image data alone indicates storage requirements on the order of a petabyte; only recently has the storage of such voluminous data become feasible as hard drive density has increased exponentially through time according to Kryder's Law [15] alongside a concomitant decrease in storage costs. The overall processing pipeline must be able to coordinate operations across different software tools and assure quality assurance through the employment of various test cases. If the data processing pipeline is implemented across sites, additional efforts will be required to coordinate the analyses. Systems to integrate data and processing pipelines across laboratories are currently being implemented, for instance, in the domain of human neuroimaging [16].

**Representation and dissemination of information**

*Neuroanatomical data model:* A major challenge is to develop an appropriate schema to represent the results of the injection experiments and associated metadata. In previous databases for connectivity information based on literature curation (CoCoMac, BAMS), the underlying model of anatomy is discrete; that is, each "connection" is associated with a pair of brain regions that are defined in the continuous brain space. Through systematic injections, and by preserving and storing original microscopy data, we can avoid such discretization and represent the results in analog form. Spatial databases [17] as used in geographical information systems provide many of the necessary tools once the underlying data model (e.g. coordinate system) has been established. Anatomical parcellations based on different atlases may then be probabilistically registered to this coordinate space to enable the representation of the full connectivity data in the form of connectivity graphs or matrices, with "nodes" defined by the particular parcellation. Furthermore, representation in the continuous space allows for a post-hoc analysis that *solves for* the partitioning of brain space that best follows the connectivity patterns observed in the data.

It is also of note that the results collected will span different levels of resolution in brain organization, from sub-neuronal compartments, neuronal compartments, neurons, groups of neurons, brain regions, to other high-level anatomical entities. The database schema should support queries designed to ask questions about various levels of organization, and should be extensible to allow incorporation of additional anatomical information, for example about cyto- or chemo-architecture or about genetic profiles of spatially localized brain areas or cell populations. The schema must support rich metadata of various forms; for the new experimental data generated within the project, the recording of all experimental parameters, dates and times, and comments from the experimenters must be explicitly built into the experimental protocols and should be as simple and automated as possible.

*Interfaces for data access and visualization:* Finally, a web-based portal should provide complete access to the data set and processed results. The "raw" data slides should be made available for fast retrieval as multi-resolution zoomable images [e.g. Ref. 18]. The primary interface for selecting and visualizing results should be a highly interactive, customizable 3D anatomical brain model (which should support rotate, pan, zoom, drag, highlight, and so on) in the spirit of tools like Google Earth [see Ref. 19]. Efficient visualization of the results is of high importance, and failures in this regard have likely been one reason for the slow adoption of database technologies in neuroscience. User interfaces will also need to support user retrieval of primary data and interoperability with other databases through Web Services and via the Semantic Web [20], leveraging emerging data protocols like XML (extensible markup language), RDF (resource description framework), and OWL Web Ontology Language.

*User groups and future efforts:* We anticipate this project will be useful to nearly all neuroscience researchers, but will be of particular interest to several groups. Because of the biomedical motivation for this project, specific interfaces to the data should be targeted toward clinical researchers and clinical trainees. Furthermore the informatics portion of the project should make every effort to integrate existing knowledge regarding the anatomical bases for neurological and neuropsychiatric disorders in animal models and through homology with humans into the knowledge base. This may best be achieved through the establishment of collaborations with medical researchers. Additional prominent user groups are expected to include theoreticians working on systems-level brain models, evolutionary neurobiologists, and researchers in functional imaging.

An important element of this coordinated collaborative effort will be to promote the use of the data set and results within the community. Furthermore, it is expected that such a large-scale project can reinvigorate and nucleate a modern neuroanatomy community. In conjunction with the suggestions outlined here, we propose the development of a new annual neuroanatomy summer course centering around these methodologies and results to help achieve such goals.

## Cost estimates and timeline

This workflow essentially describes the events that would take place at a single experimental "pipeline," of which there may be several, located at multiple sites devoted to the project. We have made coarse estimates of the equipment and personnel required to implement such experimental pipelines to carry out the proposed mouse connectivity project. The estimate is based on 5 replicates of the experimental pipeline. The major pieces of equipment required at each replicate are:

1. Mouse stereotaxic frame with motorized micromanipulator and electrode holder
2. Electrode puller
3. Current source for iontophoresis
4. Basic surgical equipment
5. Refrigeration equipment
6. Cryostat

7. Equipment for staining, immunohistochemistry, and liquids processing
8. Equipment for slide coverslipping
9. Virtual microscopy / slide scanning system

The total capital cost for one experimental "pipeline" is estimated at ~$500k USD. Assuming 5 such pipelines in distributed laboratories the estimated capital cost for experimental equipment is on the order of $2.5M. Each experiment furthermore requires certain supplies including the tracer substances and other reagents, slides and cover slips, pipettes and the mice. Here we estimate the cost for such supplies to be approximately $200 per experiment. After a startup delay, each pipeline should conservatively be capable of processing 2 experiments per day[†], or approximately 500 per year, for a recurring experimental supply cost of approximately $100k per year, per pipeline.

We assume the following personnel associated with each pipeline: 1 research assistant professor, 1 animal technician, 1 surgery / injection technician, and 1 histology / imaging technician. At an average cost of $100k per year for 4 staff members x 5 pipelines, the yearly total personnel cost for the experimental portion of the proposed program is estimated to be $2M. Additional costs will be accrued for informatics equipment and personnel, detailed below. We also summarize total expected costs and the anticipated timeline in a final section below.

*Informatics Cost Estimates:* The cost of data storage is estimated at $2 USD per gigabyte. We calculate that ~50 GB of storage is sufficient for the redundant storage of raw and processed images and metadata for each experiment, equal to $100 per experiment. Computational and server nodes plus network backbone will be a significant capital cost, approximated to be on the order of $1M plus approximately $500k for laboratory information management system (LIMS). To assess personnel costs, we assume the informatics portion of the project, which will be largely centralized, will be led by 1 research assistant professor, and will also include 2 IT technicians / programmers, and 2 postdoctoral associates. At an average of $100k annually, the IT personnel costs are on the order of $500k per year.

*Overall costs and timeline:* A summary of the proposed 5-year timeline for the mouse project is given in Figure 2. Total cost calculations are based on assumptions that the full project will require 10,000 injection experiments; given the Bota et al. (2003) estimate of 500-1000 unique cell groups (in rat), this number should be sufficient to supply multiple data points per region, and to allow for unavoidable errors. It is expected that the first project year will involve technical setup of experimental and IT infrastructure, and the experiments themselves will begin in Year 2. Thus, we estimate the need for 2,500 experiments per year, or about 10 per day. This could likely be carried out by 5 experimental work groups ("pipelines") at distributed laboratories with 4 staff members per group, as described above. Data storage costs (see above) were estimated at $100 per experiment, for a total of $1M for all experimental data. Adding capital costs for experimental equipment, virtual microscopy, and IT infrastructure plus salary for personnel (the largest expense in the project), we estimate the total cost for carrying out the proposed mouse connectivity project to be approximately $19.5M.

---

[†] Note that while the course of an individual experiment is longer than one day, we estimate that 2 animals can be taken through a particular stage of the experiment (e.g. the surgery and injection) at each pipeline in a single standard work day. This number establishes the throughput of the pipeline.

**A proposal for a coordinated effort for the determination of brainwide neuroanatomical connectivity in model organisms at a mesoscopic scale: Supporting information**

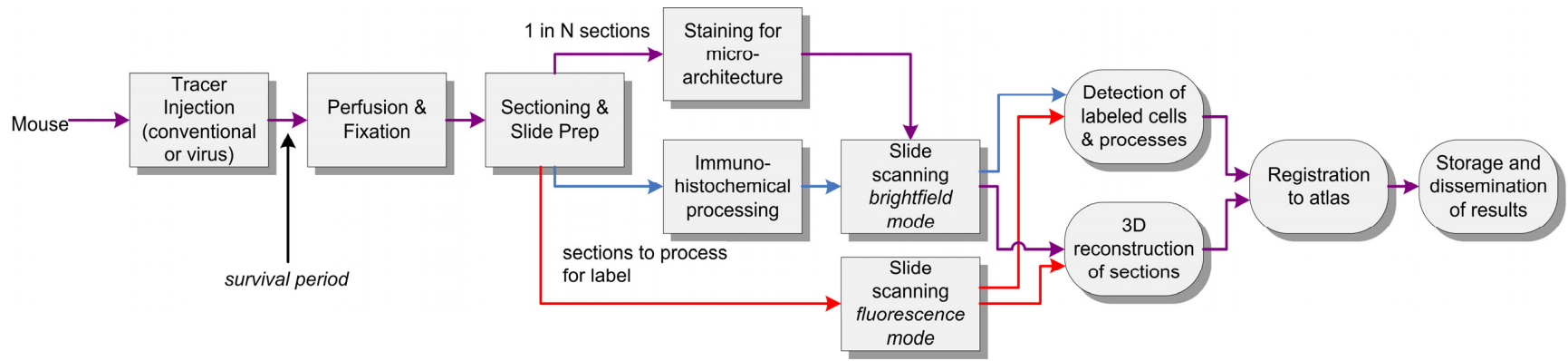

**Figure 1:** Basic workflow for a single connectivity experiment in the proposal using injections of either conventional tracers or viral vectors. Square boxes indicate stages requiring primarily experimental apparatus; rounded boxes are implemented primarily in software. Arrows represent the flow of materials (tissue, images) through the pipeline. Red arrows indicate the flow for experiments using viral tracers (and fluorescent microscopy), blue arrows indicate flow for conventional tracers, and purple arrows for flow of materials in either method.

**A proposal for a coordinated effort for the determination of brainwide neuroanatomical connectivity in model organisms at a mesoscopic scale: Supporting information**

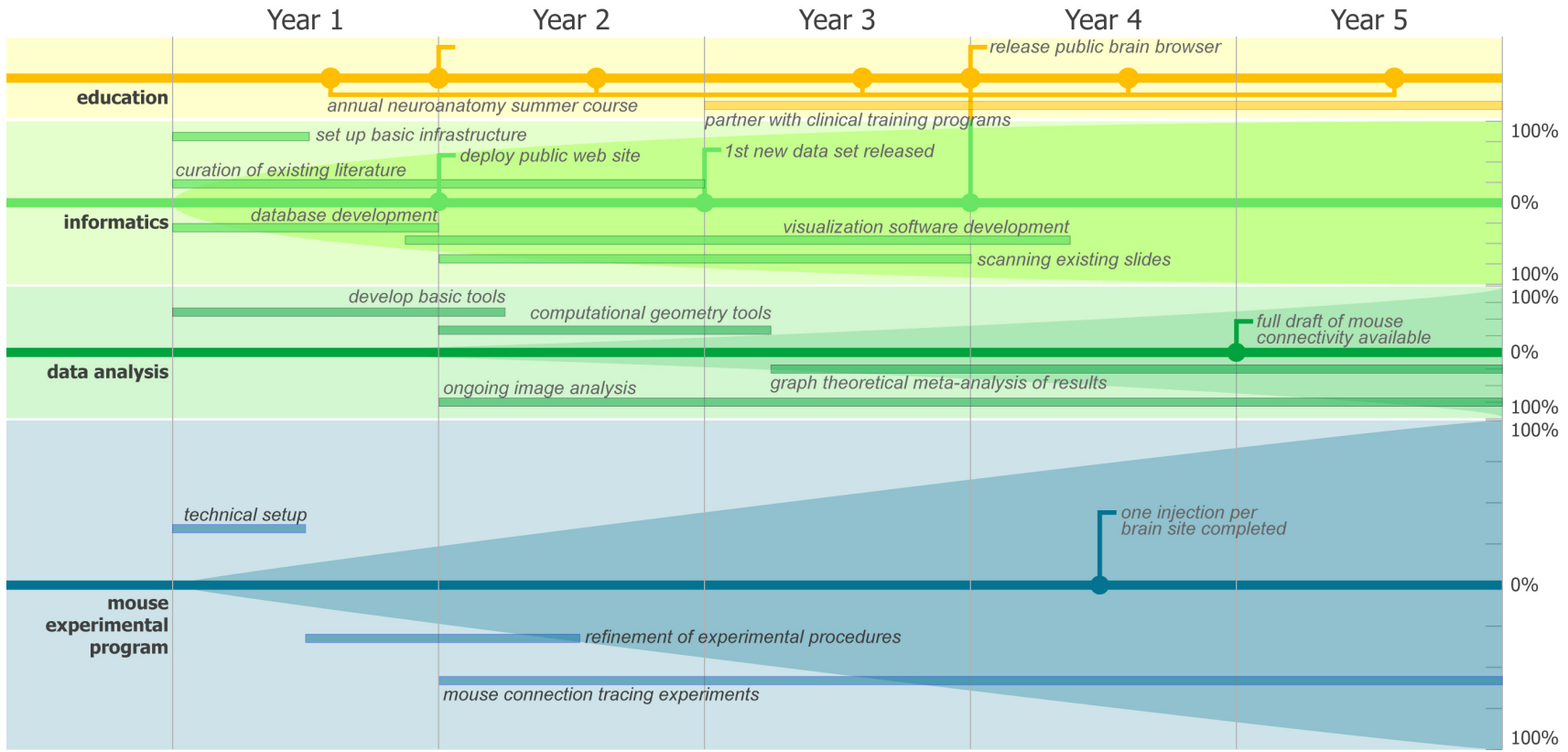

**Figure 1:** Proposed project timeline: a schematic illustration of how the proposed project would unfold in time. The project is divided into several components as depicted by the colored "bands" that run horizontally (through time). The relative thickness of each band indicates the relative distribution of project resources to that component. Events or projected milestones appear as circles on each axis, and subprojects appear as horizontal bars within each band. Finally, a set of curves in the background of each band represents the expected degree of completion of that portion of the project throughout the proposed five year span.

**A proposal for a coordinated effort for the determination of brainwide neuroanatomical connectivity in model organisms at a mesoscopic scale: Supporting information**

**A proposal for a coordinated effort for the determination of brainwide neuroanatomical connectivity in model organisms at a mesoscopic scale: Supporting information**

## Supporting Text 3: Brief proposal for primate connectivity project

As discussed in the primary article, a high-throughput investigation on the same scale as that proposed for mouse is not currently feasible in primates. While a smaller number of specifically targeted experiments in Macaque should provide highly valuable information, much effort should also be put into collating, representing, and making available the results of previous studies. Specifically this should include digitizing legacy slides, where possible, and preserving the physical slide libraries of researchers who are leaving the field. The completion of new experimental studies should then be used to "fill in" key gaps of existing knowledge or to target circuits of special importance. Such experiments should again follow standardized protocols to the extent possible (though these may differ considerably from those used in mouse) to avoid the problem of heterogeneous results that are difficult to integrate in a single global framework.

*Literature curation and capture of existing raw data:* The value of any individual connectivity study that has previously been performed in non-human primates is elevated relative to those in rodent due to the increased cost and difficulty of experimental work in primates. The CoCoMac database [1,2] currently serves an important role by gathering, through manual curation of the literature, many previous findings in macaque. While this effort should be both applauded and expanded, extraction of knowledge from articles has considerable limitations, primarily due to a lack of consistency in how results have been reported. Furthermore, textual descriptions of connections in the literature can only be considered to be an *interpretation* of the underlying data. The lack of availability of the primary source materials, beyond published images which provide only a partial view, leads to considerable uncertainty in reconciling different results. Compilations of multiple experimental results from the same laboratory using similar techniques (e.g. tritiated amino acids, HRP, BDA and other dextrans, and fluorescent dyes) provide substantial unified views of the neural connections of several cortical and subcortical structures in primates (for reviews of original papers see Refs. [3-6]). Still, a significant effort must be made to not only curate the existing literature, but to probabilistically map these results into a common 3D template space. This will help to provide an understanding of the "missing data," and will allow previous results to be viewed in the same context with targeted investigations going forward.

The collation and curation of previous results could be considerably augmented if a portion of the primary image data on which reported results were based was made available in digital form[†]. A first step in the primate portion of the project, thus, should be a concerted effort to digitize available materials in high quality using the now available technologies for virtual microscopy (see discussion of experimental methods in Supporting Text 1). Practically this could be carried out either by funding the purchase of such equipment for multiple labs that have produced large quantities of primary data, or in a centralized model in which samples are transported to a common facility for digitization and then returned to the experimentalists. Such an effort would yield a more comprehensive picture of the state of current knowledge by facilitating a more accurate *geographic* interpretation of the data (see also discussion of informatics in Supporting Text 2), rather than one based on subjective textual description.  This would require substantial computational effort to map the scanned images to a common geometric template, but such efforts could be refined over time once the primary data became digitally available.  In addition to the digital preservation of legacy slides, a central library should be created to preserve the physical slides and to allow access to future researchers.  This is particularly important as many classically trained neuroanatomists will be leaving the field in upcoming years, and their combined collections are a valuable resource that must not be discarded.  For curated and digitized legacy data, heterogeneous metadata must capture as much information about the source data as is available.  The database system should also allow the original experimentalists to annotate the legacy data by, for example, allowing "layers" of metadata (e.g. anatomical annotations) to be stored with spatial reference to the image data.

---

[†] It is unfortunately true that a considerable portion of the primary data, particularly involving fluorescent tracers, will no longer be visible due to degradation through time.

*Experimental program:* As stated above, a concerted experimental effort for macaque should be focused where current knowledge is determined to be most lacking, while also considering the potential scientific and/or biomedical impact of the new data.  For macaque studies, a multiple-labeling protocol is again proposed, though we will not describe this in detail.  In principle, the conventional tracer methods described for the mouse can be similarly applied in macaque, though a few important modifications should be made.

First, a high-quality T1-weighted anatomical magnetic resonance imaging (MRI) scan (with 1 cubic millimeter or less voxel size) should be obtained for each experimental subject.  This added preliminary stage will greatly assist in the localization of injections, in anatomical reconstruction, and in registration of individual brains to a common template.  Injection coordinates specific to each animal can be derived from analysis of the MR volume.  A second imaging data set based on diffusion spectrum imaging (DSI) [7] would have added scientific benefits if acquired in some animals, as comparison of the results of tracer injections could be compared directly with the imaging tractography in order to validate the method [see Ref. 8].  The addition of the imaging stage(s) is valuable and can be completed in a short time period, but will require the availability of imaging facilities and will add substantial cost.

The tracer protocols require additional updates and must be standardized for the macaque.  The iontophoretic injection method is not desirable for this level of investigation in the larger primate brain and should be replaced with controlled air pressure injections.  Additionally, the tracer protocol can be expanded to include several additional tracers (see, e.g. Ref. [9]) and/or multiple injection sites [10,11].  A further substantive difference will be in the post-injection survival period, which will be extended to approximately 14-18 days for the macaque. Viral methods that label multi-synaptic pathways may provide additional information in some systems, though they are more expensive and may require special laboratory equipment and expertise.  These methods are particularly viable if experiments can be implemented in existing laboratories that currently use these techniques.

Current progress being made on primate brain connectivity can be seen at BrainMaps.org [12], where sub-micron image data from recent tracer experiments using CTB, injected into various cortical areas, are directly accessible online. These ongoing efforts will result in a multi-terabyte online database of virtual slides that will enable high-resolution mapping of connectivity patterns in the primate and will drive subsequent neuroinformatics and data mining initiatives.